\documentclass[fleqn,10pt]{wlscirep}
\usepackage[utf8]{inputenc}
\usepackage[T1]{fontenc}
\usepackage{subcaption}
\title{Quantum Bohmian Inspired Potential to Model Non-Gaussian Events and the Application in Financial Markets }%

\author[1]{Reza Hosseini}
\author[2]{Samin Tajik}
\author[3]{Zahra Koohi Lai}
\author[1]{Tayeb Jamali}
\author[4]{Emmanuel Haven}
\author[1,5,*]{G. Reza Jafari}

\affil[1]{Physics Department, Shahid Beheshti University, Evin, Tehran 19839, Iran}

\affil[2]{Brock University, Physics Department, St. Catharines, ON, Canada}

\affil[3]{Department of Physics, Islamic Azad University, Firoozkooh Branch, Firoozkooh, Iran}

\affil[4]{Faculty of Business Administration, Memorial University of Newfoundland, St. John's, NL, Canada, A1C 5S7}

\affil[5]{Institute of Information Technology and Data Science, Irkutsk National Research Technical University, 83, Lermontova St., 664074 Irkutsk - Russia}

\affil[*]{g\_jafari@sbu.ac.ir}

\begin{abstract}
We have implemented quantum modeling mainly based on Bohmian Mechanics to study time series that contain strong coupling between their events. We firstly propose how compared to normal densities, our target time series seem to be associated with a higher number of rare events, and Gaussian statistics tend to underestimate these events' frequency drastically. To this end, we suggest that by imposing Gaussian densities to the natural processes, one will seriously neglect the existence of extreme events in many circumstances. The central question of our study concerns the consideration of the effects of these rare events in the corresponding probability densities and studying their role from the point of view of quantum measurements. To model the non-Gaussian behavior of these time-series, we utilize the multifractal random walk (MRW) approach and control the non-Gaussianity parameter $\lambda$ accordingly. Using the framework of quantum mechanics, we then examine the role of $\lambda$ in quantum potentials derived for these time series. Our Bohmian quantum analysis shows that the derived potential takes some negative values in high frequencies (its mean values), then substantially increases, and the value drops again for the rare events. We thus conclude that these events could generate a potential barrier that the system, lingering in a non-Gaussian high-frequency region, encounters, and their role becomes more prominent when it comes to transversing this barrier. In this study, as an example of the application of quantum potential outside of the micro-world, we compute the quantum potentials for the S\&P financial market time series to verify the presence of rare events in the non-Gaussian densities for this real data and remark the deviation from the Gaussian case.
\end{abstract}
\begin{document}

\flushbottom
\maketitle
%
%
\thispagestyle{empty}


\section*{Introduction}

From ancient times, mathematical and geometrical models have been adopted to study the world around us, and probability theories have been employed to deal with uncertainties of various events. However, in recent decades, quite some statistical experimental data in social science, notably in economics and psychology mostly within the area of human decision-making has been observed to infringe the laws of classical probability, and it has been proposed that the mathematical framework of quantum theory can offer some solutions to the challenges of this kind, as the apparatus of quantum probability theory differs significantly from the classical one.
Nowadays, we witness how societies, people, and events interact with one another on a global scale. Events happening in one corner of the globe can create a significant impact thousands of miles away. The behavior of the social and economic systems has transcended the classical framework. In the new decade, with the employment of some fundamental rules and laws from quantum theory, such as the loss of determinism, quantum superposition, and entanglement physicists aim to uncover and predict the behavior of various systems in the macro-world.
They strongly believe that in studying some of the financial and social systems, violating the laws of classical probability, a deeper uncertainty principle, relative to the uncertainty represented by classical probability theory, exists \cite{baaquie2007quantum, bohm1993book}. The number of applications from applying quantum theory to social and financial problems, ranges from cognitive science and psychology, to economy, and quantum computing for finance \cite{orus2019quantum, khrennikov2010ubiquitous, khaksar2021using, lee2020future, bagarello2012quantum, mugel2021hybrid}. In finance, more specifically, quantum modeling of risks and decision-making to the analysis of the financial market \cite{woerner2019quantum, tahmasebi2015financial, choustova2009quantum, chabaud1994b}, has concerned interest rates and option pricing \cite{haven2003h, nasiri2018impact, haven2013quantum}. According to these studies, the classical price dynamics can no longer be applied to all modern financial markets to study the price trajectories of these markets, and one needs to also consider the significant role played by multiple behavioral factors. Moreover,the fact that traders in modern financial markets tend to behave stochastically given their free wills is needed to be taken into account. One of the most prominent deficiencies of applying classical mechanics to uncover behaviors of financial markets reveals itself in the non-locality-like features of the modern financial markets. In order to study modern financial markets we need to consider the price return of a period, consisting of several days "entangled" to each other. The quantum mechanical approach uses extended ideas based on quantum entanglement to examine the correlation of different time series with a high number of extreme events to study and predict their evolution in time. In this paper, we have employed the Bohmian quantum potential method in order to study an example of these entanglements in financial markets and analyze the impacts of extreme events on these time series. 

\section*{Bohmian Mechanics and Quantum Potential Inspired Method}

In this section we show how the Schrodinger wave function which is the heart of quantum mechanics can be used to explain the relationship between events in space-time, and accordingly to address the obstacle of non-locality of events. The notion that the wave function of a system, evolving according to the Schrodinger equation, is interpreted as an active information field, builds the foundation of Bohmian mechanics, in which information, at the level of human perception, functions according to postulates from information at the quantum level. This approach is fundamentally based on studies of D. Bohm, B. Hiley, and P. Pylkkänen,  on the active information analysis of Bohmian Mechanics and its applications to cognitive sciences \cite{bohm1952suggested, choustova2007quantum}. Following these studies, and also previous work done by E. Haven, A Khrennikov, C. Shen and G. Jafari  \cite{haven2017applications, shen2017using}, we pursued Schrodinger formalism to demonstrate how Bohmian mechanics complies with our understanding of financial markets as an example of a correlated system. We represent our financial pilot-wave $\psi(q)$, evolving according to Schrodinger's equation in the following form:

\begin{equation}
        \psi(q,t) = R(q,t)e^{iS(q,t)/\hbar},
\end{equation}
    
where $R(q,t) = |\psi(q,t)|$ is the amplitude, and $S(q,t)$ is the phase of the defined wave function $\psi(q,t)$.
Substituting the pilot wave into the Schrodinger equation yields:

\begin{equation}
        \frac{\partial R^2}{\partial t}+\frac{1}{m}\frac{\partial(R^2 \frac{\partial S}{\partial q})}{\partial q}=0,
    \end{equation}
    \begin{equation}
        \frac{\partial S}{\partial t}+\frac{1}{2m}(\frac{\partial S}{\partial q})^2+(V-U)=0,
\end{equation}
    
where the Bohm quantum potential is:

\begin{equation}
        U=\frac{\hbar^2}{2mR}\frac{\partial^2 R}{\partial q^2}.
\end{equation}

In the following section, we demonstrate how we want to analyze the dynamics of non-Gaussian functions with the help of the Bohmian quantum potential. We first give a review on the multifractal process and multifractal random walk to model the non-Gaussian probability density function of our desired data.

\section*{Multifractal Formalism}

Multifractality has been previously applied to consider scale invariance features of various objectives in several areas of research. Different studies have examined the concept of pairing multifractality between time series based on the increasing number of rare events that can generate deviation from the gaussian density~\cite{lai2015coupled, ausloos2012generalized,IJMPC2007}. In particular, the MRW models, popular in modeling stock fluctuations in the financial market, have become the focus of some recent analyses~\cite{saakian2011exact, muzy2001multifractal}. Moreover, the relationship between turbulence and finance has triggered the demand for employing multifractal models which previously studied in the framework of multiplicative random cascades \cite{muzy2000modelling, ghashghaie1996turbulent}.  Here we aim to employ multifractal concepts to account for scale invariant properties of financial data, based on which the robust technique of MRW is introduced and applied \cite{bacry2001multifractal}.

The non-Gaussian probability density function (PDF) with the robust multifractality arises from the strong log-normal deviation from the normal state which primarily is due to the occurrence of large fluctuations in the data set. The exact multifractal properties are a consequence of the correlation in the argument of the logarithm of the stochastic variances \cite{kiyono2006criticality}.

Consider a stochastic process, represented by $X(t)$, which may be a function of space-time, the increment fluctuations of the data sets at a time scale $\tau$ can be shown as:
    \begin{equation}
        \Delta X_{\tau}(t)=X(t+\tau)-X(t).
    \end{equation}
    The process is called scale-invariant when the absolute moment M(q), has the following power-law behavior:
    \begin{equation}
        M(q,\tau)=\left< |\Delta X (\tau)|^q\right> = M(q,\tau)\propto \tau^{\xi_q},
    \end{equation}
    where we define $\xi_q$ as the exponent of the power law, and is responsible for characterizing the scale invariance properties of the fractal function, \cite{manshour2009turbulencelike} and can be shown by:
    
    \begin{equation}
        \xi_q= qH-1/2 (q(q-2)\lambda^2).
        \label{xi}
    \end{equation}
    
The process is then called monofractal if $\xi_q$ is a linear function of $q$, and multifractal if $\xi_q$ is a nonlinear function of q \cite{shayeganfar2009multifractal}. 
In Eq. \ref{xi}, $H$ is the Hurst scaling exponent of time series, such that $0<H<1$. For $0<H<0.5$ the system is known to be anti correlated, $0.5<H<1$ leads to correlation and for $H=0$ we have an uncorrelated system. The value of $\lambda$ scales the non Gasussianity, such that for the Gaussian case we have: $\lambda=0$, corresponding to $\xi_q \propto q$, and indicating the fractality of the signal. For the non Gaussian case with the $\lambda \neq 0$ signal will represent multifractality behavior. In this section, using only a set of few variables, we apply multifractal statistics to model the increment of fluctuations of the data \cite{koohi2021coupled, shayeganfar2010stochastic}.
    
For a log-normal cascade at the smallest scale $\Delta t$ and for each time-lag, $\tau$, to obtain a good candidate that satisfies the cascading relation, known as the MRW, we write:
    \begin{equation}
        \Delta_{\tau} X(t)=  \epsilon(t) e^{\omega(t)},
    \end{equation}
    
where $\epsilon(t)$, and $\omega(t)$ are Gaussian variables with their corresponding variances being indicated by $\sigma^2$, and $\lambda^2$, respectively. In this approach, regarding the stated stochastic variables, in order to convey the analytical calculation of Quantum potential for the stated model we first need to define the non-Gaussian probability density function (PDF) with fat tails time series. Hence, we can find the relationship between non-Gaussian parameters, $\lambda$ and multifratality which comes from the nonlinear function general exponents $\xi_q$ vs $q$. Based on the Castaing model, a process is called self-similar if the increment's probability density functions at scales $\tau$ are related by the following equation\cite{castaing1990velocity, chabaud1994b}
    
    \begin{equation}
        P(\Delta_{\tau} X)= \int G_{\tau}(Ln\sigma) \frac{1}{\sigma}F_{\tau} \left(\frac{\Delta_l x}{\sigma}\right) d ln\sigma,
        \label{pdf}
    \end{equation}
    where
    \begin{equation}
        G(Ln\sigma)=\frac{1}{\sqrt{2\pi}\lambda}exp\left(-\frac{ln^2\sigma}{2\lambda^2}\right)
    \end{equation}

\begin{equation}
        F(\frac{\Delta_\tau x}{\sigma})=\frac{1}{\sqrt{2\pi}}exp\left(-\frac{(\Delta_\tau x)^2}{2\sigma^2}\right).
    \end{equation}

\begin{figure}[h]
	\centering
	\begin{minipage}{.55\columnwidth}
		\centering
		\includegraphics[width=0.9\textwidth]{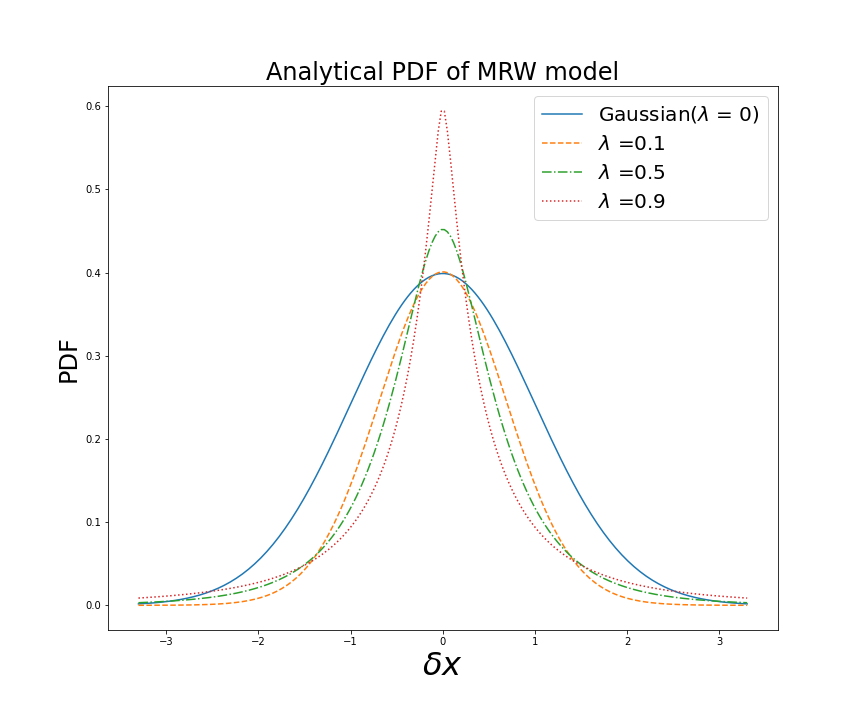}
		\caption*{(a)}
		\label{PD0}
		
	\end{minipage}%
	\begin{minipage}{.55\columnwidth}
		\centering
		\includegraphics[width=0.9\textwidth]{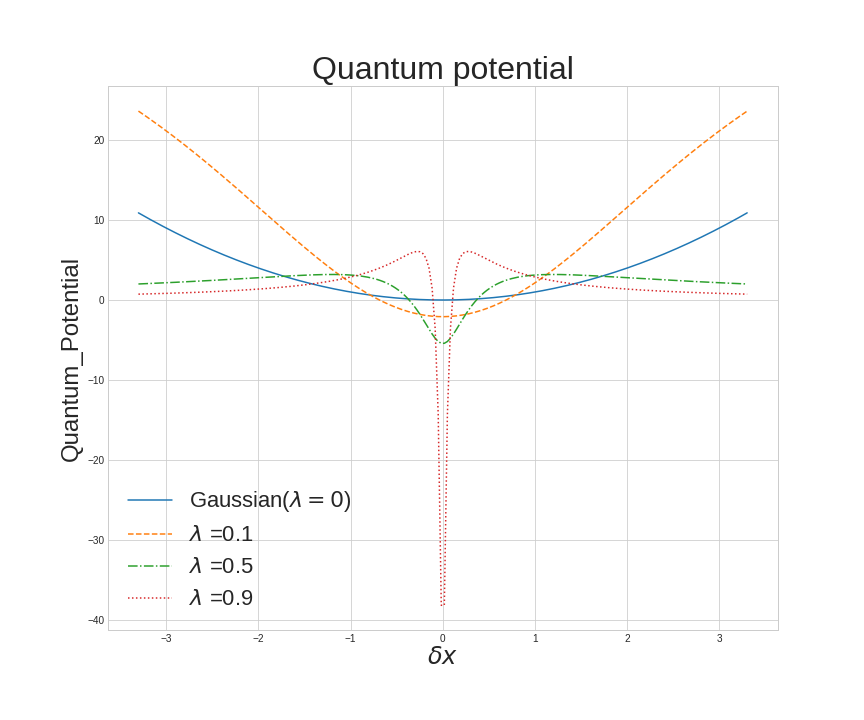}
		\caption*{(b)}
		\label{betti0}
		
	\end{minipage}
	\caption{a. The density function for the cascade model with three different non-Gaussian parameter values $\lambda \in [0.1,0.5,0.9]$
            and $\sigma =1$. We also compare it with Gaussian density function with $\sigma =1$. b. The figure shows the corresponding derived Quantum potential.}
	\label{pdfqps}
\end{figure}

The Bohmian quantum potential depends only on the second spatial derivative of the amplitude of the wave, taking $P = R^2$, we write: 

    \begin{equation}
       \frac{d^2P(x)}{dx^2}=\frac{1}{\pi\lambda}\int_{0}^\infty{\frac
            {(2x^2-\sigma^2)}{\sigma^5}e^{-\frac{\ln^2{\frac{\sigma}{\sigma_0}}}{2\lambda^2}}e^{-\frac{x^2}{\sigma^2}}} (d ln\sigma).
    \end{equation}

We now compute the quantum potential ($U \propto \frac{1}{P(x)} \frac{d^2P(x)}{dx^2}$) for the above probability density function (without considering power two of R) and plot them for a range of parameters for further comparison.

\section*{Results for Computing Quantum Potential and Real Data Fit}

\begin{figure}[h]
	\includegraphics[width=0.5 \textwidth,]{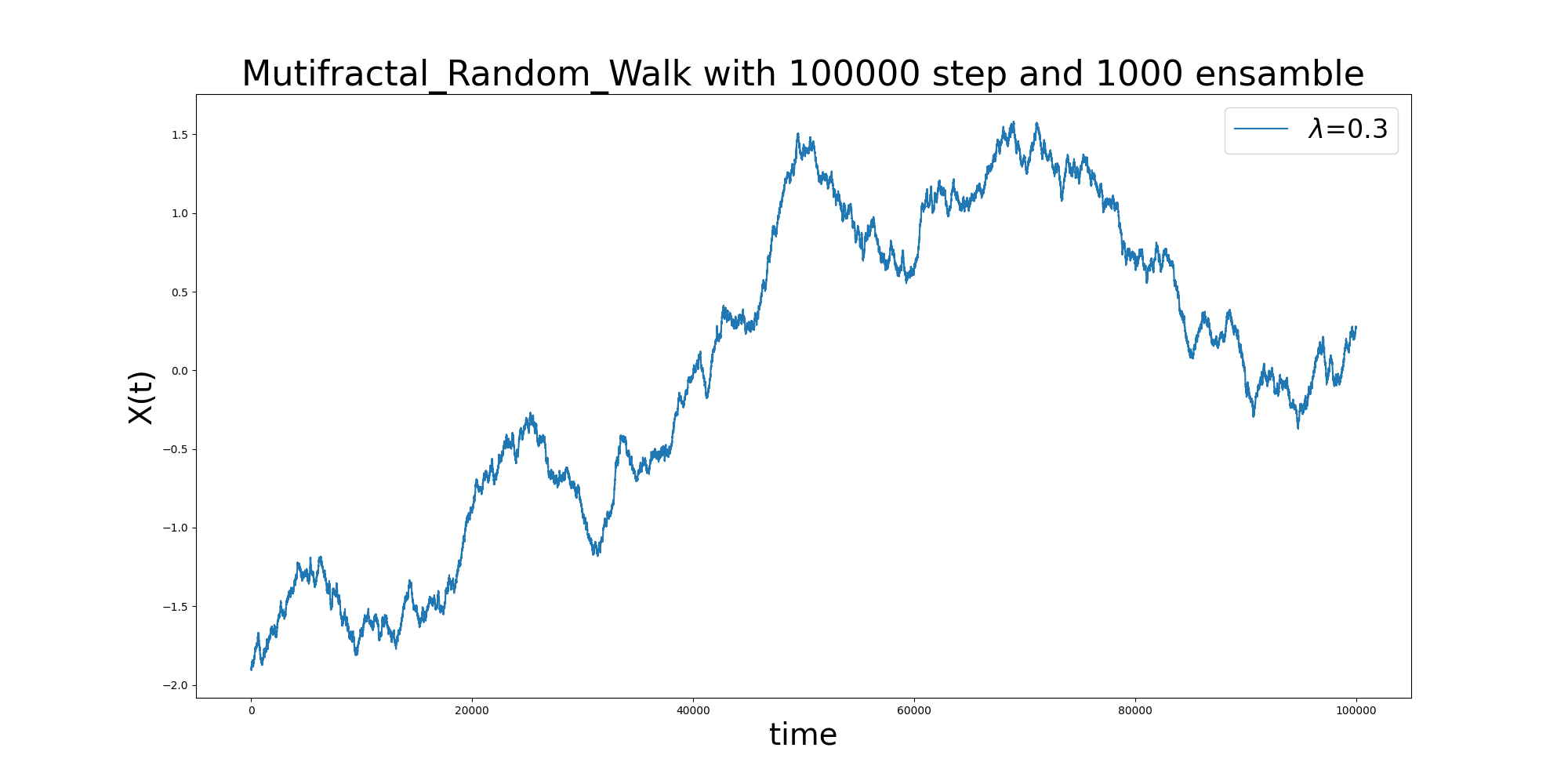}
	\includegraphics[width=0.45 \textwidth,]{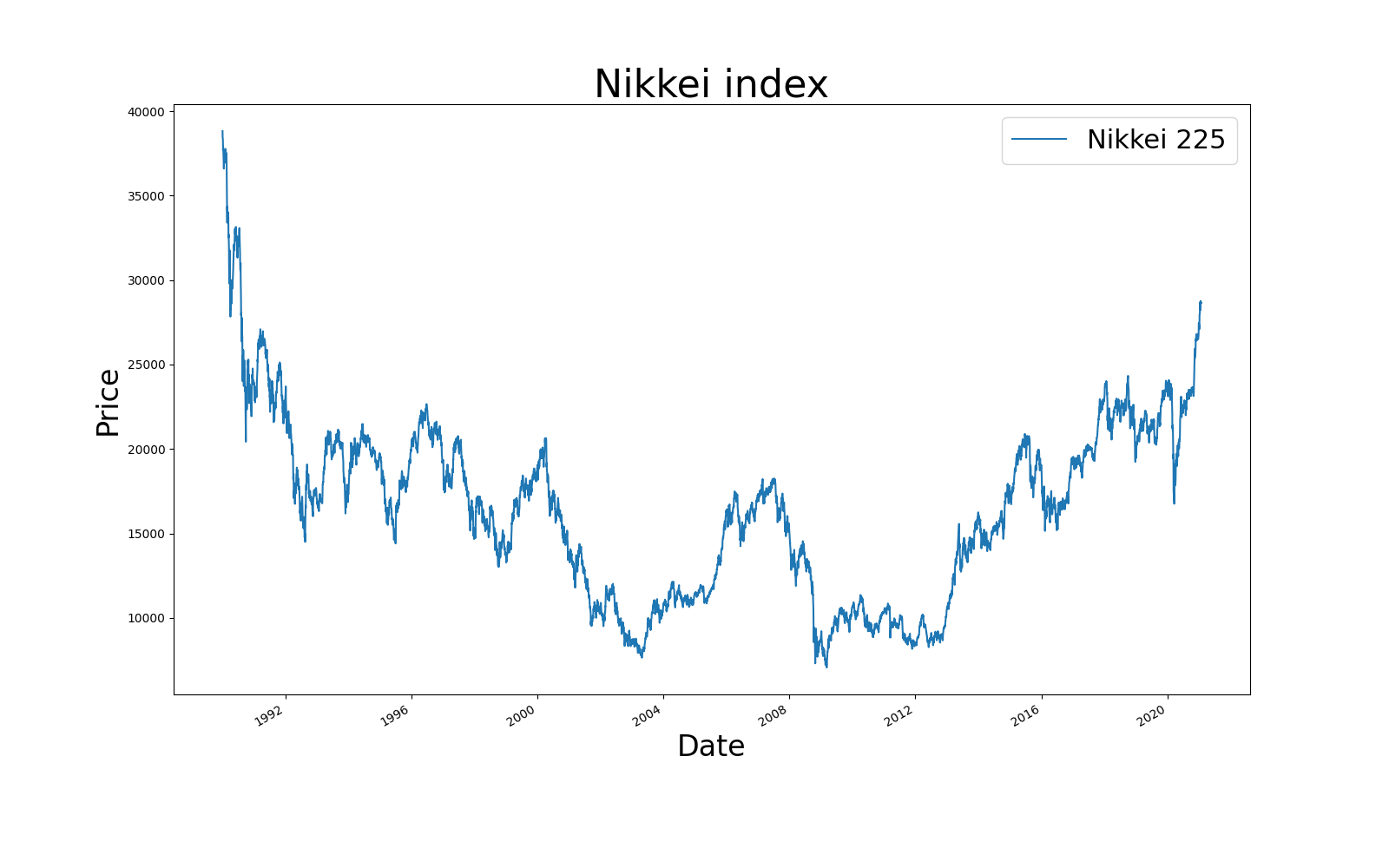}
	\includegraphics[width=0.5 \textwidth,]{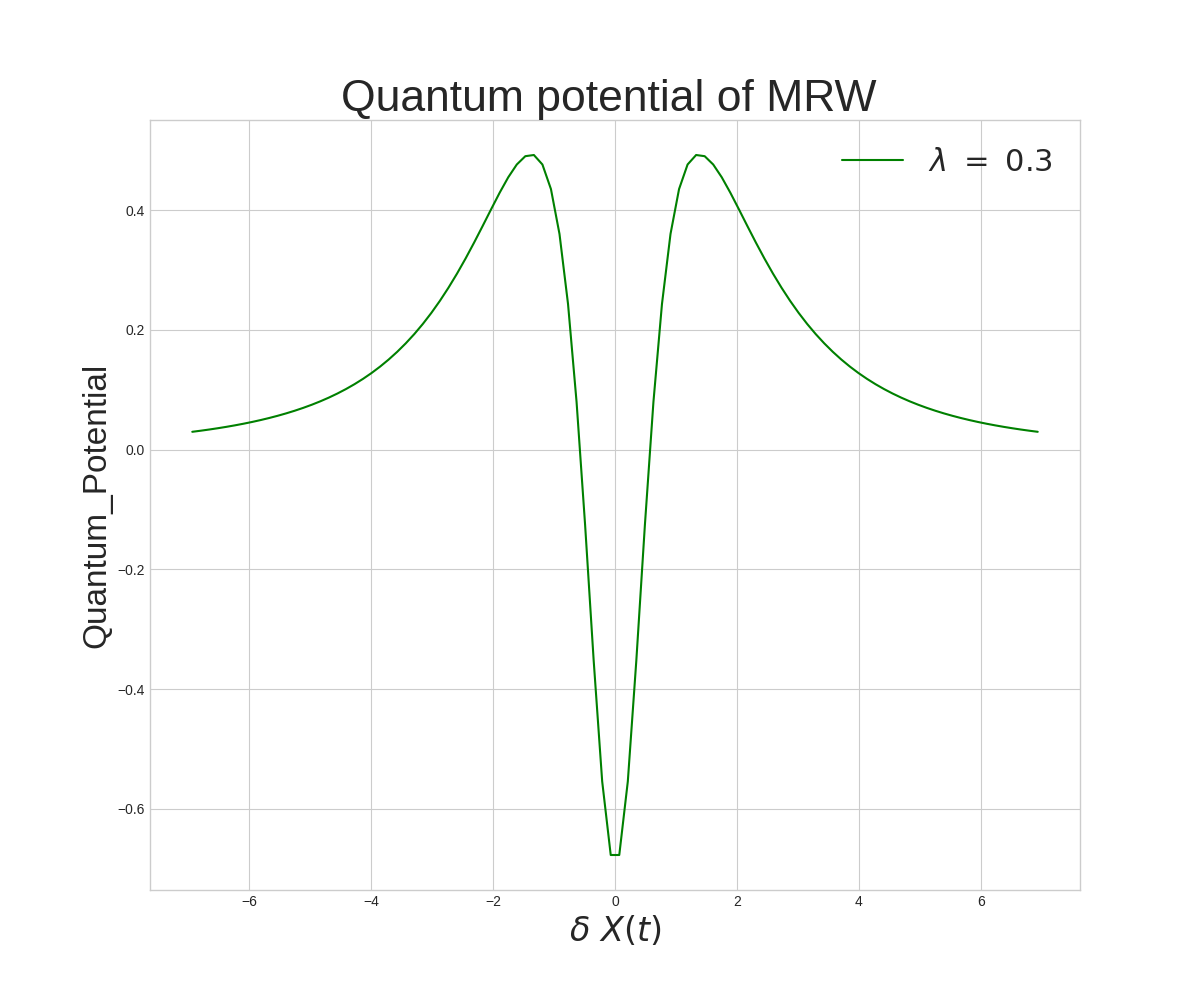}
	\includegraphics[width=0.5 \textwidth,]{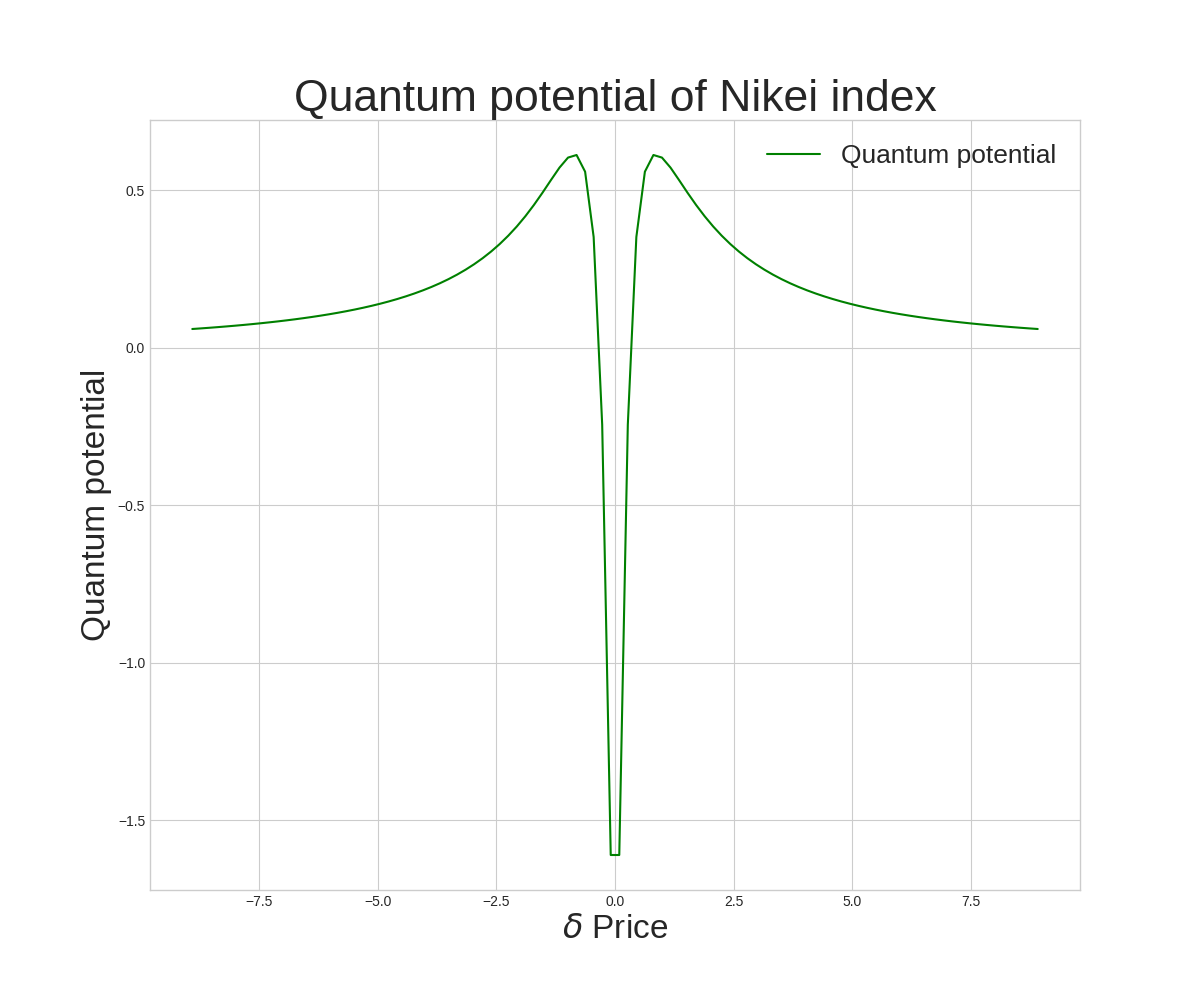}
	\caption{MRW implemented with $100000$ steps and $\sigma_0 = 1$, in $\lambda=0.3$, $H=0.6$, and $L=2000$, its return, and corresponding quantum potential, together Nikkei index data from 1990 to 2020 and its normalized return, bottom panel shows the corresponding quantum potential for the toy model(left panel), and for Nikkei(right panel).}
	\label{mrw-nikkei}
\end{figure}

Based on our previous arguments, employing Bohmian mechanics and studying the strong financial effects on the market trajectories, we aim to describe the dynamics of the financial pilot wave.
With the use of probability density function defined as Eq. \ref{pdf} for non-Gaussian functions with a range of $\lambda$'s, we examine the effects of extreme events, and compute their corresponding quantum potentials to analyze and fit real market data subsequently. In our calculation we consider $\hbar$ as a price scaling parameter and will assume $\hbar=1$. In Fig.\ref{pdfqps} we compare the contribution of rare events on the quantum potential for Gaussian density $(\lambda=0)$ and non-Gaussian density functions with several $\lambda's$.
The left panel of Fig\ref{pdfqps} shows some density functions for $\sigma=1$ and a range of $\lambda$s, and the right panel displays the corresponding calculated quantum potentials.
As it can be noted from the graph, increasing the $\lambda$ to leads to a remarkable distortion promoting the occurrence of the extreme events. Moreover, this figure shows how this increase in the value of $\lambda$ will induce a potential barrier, which when transcended will provoke the system to proceed to its critical state.
Our quantum potential analysis shows there exist two main contributions to this potential. Firstly, one can discuss the vicinity of mid-value where the system holds a low quantum potential, revealing the system's tendency to survive in this region. Upon increasing the non-Gaussianity parameter we remarked that the quantum potential which would increase and act as a potential well is now shrinking. This reveals its contribution to the occurrence of the influential events in the system. However, with the further increase in the non-Gaussian parameter, we observed that the quantum potential for these big events compared to the mean is reduced. The presence of these two phases uncovers the aim of the system to stay in the region where extreme events take place.
As was expected, the behavior of a quantum barrier, revealing the limitation that the price range faces, explains how increasing the non-Gaussian parameter $\lambda$, in spite of the rise in the depth of the well, does not induce the infinity in the well. From this behavior we concluded that any market whose energy is large enough to surpass this barrier will enter its critical phase.
Next, following the multifractal random walk formalism we construct a toy model with $\lambda=0.3$ and $\sigma=1$, as shown in Fig.\ref{mrw-nikkei}, and calculate the corresponding quantum potential for our model. As can be remarked from the graph, for this $\lambda$ the energy of the model exceeds the potential barrier, and consequently, the system undergoes its critical condition. To analyze the real market and compare it to our toy model we also have illustrated the behavior of the Nikkei index from 1990 to 2020 in the same figure and examined it. The market PDF, with $\sigma = 0.802$ shows how it sustains its critical stage with the average $\lambda$ of $0.399$. As we noted earlier, the analysis of quantum potential of the Nikkei index unveils that compared to Gaussian density functions, the density functions of these markets contain plenty of sudden crises in this era. The main difference between our model and the real data manifests itself in the value for the return. As denoted from the Fig.\ref{mrw-nikkei} unlike the behavior of our constructed toy model, the real market return function tends to behave inhomogeneously. We believe that this deviation comes from the fact that in our model $\lambda$ is set to a uniform value. Nonetheless, real markets with the same average value of $\lambda$ sustain a relatively nonstationary character. Considering the inconstancies, $\lambda$ in these series would have some time dependency, and therefore we require to consider an average value for our analysis.
\section*{Conclusion}

At present, various research studies have been utilizing models and techniques to apply the laws of a quantum system to the macro world's social systems and financial markets. Moreover, one of the most distinctive features of financial markets' data lies in their non-Gaussian performance and the probability of extreme events occurring in the markets. Upon the question of whether this behavior stems from an underlying structural pattern, we employed an innovative method to analyze these functions in the framework of quantum theory. Thus, by implementing quantum Bohmian mechanics and deriving the corresponding quantum potentials for the non-Gaussian density function of the data, we can identify the distinction between the behavior of extreme events in the financial markets explained by Gaussian and non-Gaussian functions. Besides, we confirmed how these extreme events in the non-Gaussian cases introduce themselves via a potential barrier in a process, aiming to linger in its mean values. However, we cannot oversee the fact that the occurrence of extreme events has a significant role in these processes. Consequently, in this work, we verified the application of using the quantum potential for non-Gaussian functions to clarify the behavior of financial markets. Nevertheless, we firmly believe that there is a wide range of underlying applications to this approach for a variety of natural processes.

\bibliography{sample}

\begin{thebibliography}{10}
\urlstyle{rm}
\expandafter\ifx\csname url\endcsname\relax
  \def\url#1{\texttt{#1}}\fi
\expandafter\ifx\csname urlprefix\endcsname\relax\def\urlprefix{URL }\fi
\expandafter\ifx\csname doiprefix\endcsname\relax\def\doiprefix{DOI: }\fi
\providecommand{\bibinfo}[2]{#2}
\providecommand{\eprint}[2][]{\url{#2}}

\bibitem{baaquie2007quantum}
\bibinfo{author}{Baaquie, B.~E.}
\newblock \emph{\bibinfo{title}{Quantum finance: Path integrals and
  Hamiltonians for options and interest rates}} (\bibinfo{publisher}{Cambridge
  University Press}, \bibinfo{year}{2007}).

\bibitem{bohm1993book}
\bibinfo{author}{Bohm, D.}, \bibinfo{author}{Hiley, B.} \&
  \bibinfo{author}{Holland, P.}
\newblock \bibinfo{journal}{\bibinfo{title}{Book-review-the undivided
  universe-an ontological interpretation of quantum theory}}.
\newblock {\emph{\JournalTitle{Nature}}} \textbf{\bibinfo{volume}{366}},
  \bibinfo{pages}{420} (\bibinfo{year}{1993}).

\bibitem{orus2019quantum}
\bibinfo{author}{Orus, R.}, \bibinfo{author}{Mugel, S.} \&
  \bibinfo{author}{Lizaso, E.}
\newblock \bibinfo{journal}{\bibinfo{title}{Quantum computing for finance:
  Overview and prospects}}.
\newblock {\emph{\JournalTitle{Reviews in Physics}}}
  \textbf{\bibinfo{volume}{4}}, \bibinfo{pages}{100028} (\bibinfo{year}{2019}).

\bibitem{khrennikov2010ubiquitous}
\bibinfo{author}{Khrennikov, A.}
\newblock \emph{\bibinfo{title}{Ubiquitous quantum structure}}
  (\bibinfo{publisher}{Springer}, \bibinfo{year}{2010}).

\bibitem{khaksar2021using}
\bibinfo{author}{Khaksar, H.}, \bibinfo{author}{Haven, E.},
  \bibinfo{author}{Nasiri, S.} \& \bibinfo{author}{Jafari, G.}
\newblock \bibinfo{journal}{\bibinfo{title}{Using the quantum potential in
  elementary portfolio management: Some initial ideas}}.
\newblock {\emph{\JournalTitle{Entropy}}} \textbf{\bibinfo{volume}{23}},
  \bibinfo{pages}{180} (\bibinfo{year}{2021}).

\bibitem{lee2020future}
\bibinfo{author}{Lee, R.~S.}
\newblock \bibinfo{title}{Future trends in quantum finance}.
\newblock In \emph{\bibinfo{booktitle}{Quantum Finance}},
  \bibinfo{pages}{399--405} (\bibinfo{publisher}{Springer},
  \bibinfo{year}{2020}).

\bibitem{bagarello2012quantum}
\bibinfo{author}{Bagarello, F.}
\newblock \emph{\bibinfo{title}{Quantum dynamics for classical systems: with
  applications of the number operator}} (\bibinfo{publisher}{John Wiley \&
  Sons}, \bibinfo{year}{2012}).

\bibitem{mugel2021hybrid}
\bibinfo{author}{Mugel, S.} \emph{et~al.}
\newblock \bibinfo{journal}{\bibinfo{title}{Hybrid quantum investment
  optimization with minimal holding period}}.
\newblock {\emph{\JournalTitle{Scientific Reports}}}
  \textbf{\bibinfo{volume}{11}}, \bibinfo{pages}{1--6} (\bibinfo{year}{2021}).

\bibitem{woerner2019quantum}
\bibinfo{author}{Woerner, S.} \& \bibinfo{author}{Egger, D.~J.}
\newblock \bibinfo{journal}{\bibinfo{title}{Quantum risk analysis}}.
\newblock {\emph{\JournalTitle{npj Quantum Information}}}
  \textbf{\bibinfo{volume}{5}}, \bibinfo{pages}{1--8} (\bibinfo{year}{2019}).

\bibitem{tahmasebi2015financial}
\bibinfo{author}{Tahmasebi, F.} \emph{et~al.}
\newblock \bibinfo{journal}{\bibinfo{title}{Financial market images: a
  practical approach owing to the secret quantum potential}}.
\newblock {\emph{\JournalTitle{EPL (Europhysics Letters)}}}
  \textbf{\bibinfo{volume}{109}}, \bibinfo{pages}{30001}
  (\bibinfo{year}{2015}).

\bibitem{choustova2009quantum}
\bibinfo{author}{Choustova, O.}
\newblock \bibinfo{journal}{\bibinfo{title}{Quantum probability and financial
  market}}.
\newblock {\emph{\JournalTitle{Information Sciences}}}
  \textbf{\bibinfo{volume}{179}}, \bibinfo{pages}{478--484}
  (\bibinfo{year}{2009}).

\bibitem{chabaud1994b}
\bibinfo{author}{Chabaud, B.}, \bibinfo{author}{Naert, A.},
  \bibinfo{author}{Peinke, J.} \& \bibinfo{author}{Chilla, F.}
\newblock \bibinfo{journal}{\bibinfo{title}{, b. castaing, and b. h{\'e}bral}}.
\newblock {\emph{\JournalTitle{Phys. Rev. Lett}}}
  \textbf{\bibinfo{volume}{73}}, \bibinfo{pages}{3227} (\bibinfo{year}{1994}).

\bibitem{haven2003h}
\bibinfo{author}{Haven, E.}
\newblock \bibinfo{journal}{\bibinfo{title}{An ‘h-brownian motion’of
  stochastic option prices}}.
\newblock {\emph{\JournalTitle{Physica A}}} \textbf{\bibinfo{volume}{344}},
  \bibinfo{pages}{151--155} (\bibinfo{year}{2003}).

\bibitem{nasiri2018impact}
\bibinfo{author}{Nasiri, S.}, \bibinfo{author}{Bektas, E.} \&
  \bibinfo{author}{Jafari, G.~R.}
\newblock \bibinfo{journal}{\bibinfo{title}{The impact of trading volume on the
  stock market credibility: Bohmian quantum potential approach}}.
\newblock {\emph{\JournalTitle{Physica A: Statistical Mechanics and its
  Applications}}} \textbf{\bibinfo{volume}{512}}, \bibinfo{pages}{1104--1112}
  (\bibinfo{year}{2018}).

\bibitem{haven2013quantum}
\bibinfo{author}{Haven, E.} \& \bibinfo{author}{Khrennikov, A.}
\newblock \emph{\bibinfo{title}{Quantum social science}}.

\bibitem{bohm1952suggested}
\bibinfo{author}{Bohm, D.}
\newblock \bibinfo{journal}{\bibinfo{title}{A suggested interpretation of the
  quantum theory in terms of" hidden" variables. i}}.
\newblock {\emph{\JournalTitle{Physical review}}}
  \textbf{\bibinfo{volume}{85}}, \bibinfo{pages}{166} (\bibinfo{year}{1952}).

\bibitem{choustova2007quantum}
\bibinfo{author}{Choustova, O.}
\newblock \bibinfo{journal}{\bibinfo{title}{Quantum modeling of nonlinear
  dynamics of stock prices: Bohmian approach}}.
\newblock {\emph{\JournalTitle{Theoretical and Mathematical Physics}}}
  \textbf{\bibinfo{volume}{152}}, \bibinfo{pages}{1213--1222}
  (\bibinfo{year}{2007}).

\bibitem{haven2017applications}
\bibinfo{author}{Haven, E.} \& \bibinfo{author}{Khrennikov, A.}
\newblock \bibinfo{journal}{\bibinfo{title}{Applications of quantum mechanical
  techniques to areas outside of quantum mechanics}}.
\newblock {\emph{\JournalTitle{Frontiers in Physics}}}
  \textbf{\bibinfo{volume}{5}}, \bibinfo{pages}{60} (\bibinfo{year}{2017}).

\bibitem{shen2017using}
\bibinfo{author}{Shen, C.} \& \bibinfo{author}{Haven, E.}
\newblock \bibinfo{journal}{\bibinfo{title}{Using empirical data to estimate
  potential functions in commodity markets: Some initial results}}.
\newblock {\emph{\JournalTitle{International Journal of Theoretical Physics}}}
  \textbf{\bibinfo{volume}{56}}, \bibinfo{pages}{4092--4104}
  (\bibinfo{year}{2017}).

\bibitem{lai2015coupled}
\bibinfo{author}{Lai, Z.~K.}, \bibinfo{author}{Farahani, S.~V.},
  \bibinfo{author}{Movahed, S.} \& \bibinfo{author}{Jafari, G.}
\newblock \bibinfo{journal}{\bibinfo{title}{Coupled uncertainty provided by a
  multifractal random walker}}.
\newblock {\emph{\JournalTitle{Physics Letters A}}}
  \textbf{\bibinfo{volume}{379}}, \bibinfo{pages}{2284--2290}
  (\bibinfo{year}{2015}).

\bibitem{ausloos2012generalized}
\bibinfo{author}{Ausloos, M.}
\newblock \bibinfo{journal}{\bibinfo{title}{Generalized hurst exponent and
  multifractal function of original and translated texts mapped into frequency
  and length time series}}.
\newblock {\emph{\JournalTitle{Physical Review E}}}
  \textbf{\bibinfo{volume}{86}}, \bibinfo{pages}{031108}
  (\bibinfo{year}{2012}).

\bibitem{IJMPC2007}
\bibinfo{author}{G.R.~Jafari, P. N. A. B. M. S. F.~G., M. Sadegh~Movahed} \&
  \bibinfo{author}{Tabar, M. R.~R.}
\newblock \bibinfo{journal}{\bibinfo{title}{Uncertainty in the fluctuations of
  the price of stocks}}.
\newblock {\emph{\JournalTitle{International Journal of Modern Physics C}}}
  \textbf{\bibinfo{volume}{18}}, \bibinfo{pages}{1689} (\bibinfo{year}{2007}).

\bibitem{saakian2011exact}
\bibinfo{author}{Saakian, D.~B.}, \bibinfo{author}{Martirosyan, A.},
  \bibinfo{author}{Hu, C.-K.} \& \bibinfo{author}{Struzik, Z.}
\newblock \bibinfo{journal}{\bibinfo{title}{Exact probability distribution
  function for multifractal random walk models of stocks}}.
\newblock {\emph{\JournalTitle{EPL (Europhysics Letters)}}}
  \textbf{\bibinfo{volume}{95}}, \bibinfo{pages}{28007} (\bibinfo{year}{2011}).

\bibitem{muzy2001multifractal}
\bibinfo{author}{Muzy, J.-F.}, \bibinfo{author}{Sornette, D.},
  \bibinfo{author}{Delour, J.} \& \bibinfo{author}{Arneodo, A.}
\newblock \bibinfo{journal}{\bibinfo{title}{Multifractal returns and
  hierarchical portfolio theory}}.
\newblock  (\bibinfo{year}{2001}).

\bibitem{muzy2000modelling}
\bibinfo{author}{Muzy, J.-F.}, \bibinfo{author}{Delour, J.} \&
  \bibinfo{author}{Bacry, E.}
\newblock \bibinfo{journal}{\bibinfo{title}{Modelling fluctuations of financial
  time series: from cascade process to stochastic volatility model}}.
\newblock {\emph{\JournalTitle{The European Physical Journal B-Condensed Matter
  and Complex Systems}}} \textbf{\bibinfo{volume}{17}},
  \bibinfo{pages}{537--548} (\bibinfo{year}{2000}).

\bibitem{ghashghaie1996turbulent}
\bibinfo{author}{Ghashghaie, S.}, \bibinfo{author}{Breymann, W.},
  \bibinfo{author}{Peinke, J.}, \bibinfo{author}{Talkner, P.} \&
  \bibinfo{author}{Dodge, Y.}
\newblock \bibinfo{journal}{\bibinfo{title}{Turbulent cascades in foreign
  exchange markets}}.
\newblock {\emph{\JournalTitle{Nature}}} \textbf{\bibinfo{volume}{381}},
  \bibinfo{pages}{767--770} (\bibinfo{year}{1996}).

\bibitem{bacry2001multifractal}
\bibinfo{author}{Bacry, E.}, \bibinfo{author}{Delour, J.} \&
  \bibinfo{author}{Muzy, J.-F.}
\newblock \bibinfo{journal}{\bibinfo{title}{Multifractal random walk}}.
\newblock {\emph{\JournalTitle{Physical Review E}}}
  \textbf{\bibinfo{volume}{64}}, \bibinfo{pages}{026103}
  (\bibinfo{year}{2001}).

\bibitem{kiyono2006criticality}
\bibinfo{author}{Kiyono, K.}, \bibinfo{author}{Struzik, Z.~R.} \&
  \bibinfo{author}{Yamamoto, Y.}
\newblock \bibinfo{journal}{\bibinfo{title}{Criticality and phase transition in
  stock-price fluctuations}}.
\newblock {\emph{\JournalTitle{Physical review letters}}}
  \textbf{\bibinfo{volume}{96}}, \bibinfo{pages}{068701}
  (\bibinfo{year}{2006}).

\bibitem{manshour2009turbulencelike}
\bibinfo{author}{Manshour, P.} \emph{et~al.}
\newblock \bibinfo{journal}{\bibinfo{title}{Turbulencelike behavior of seismic
  time series}}.
\newblock {\emph{\JournalTitle{Physical review letters}}}
  \textbf{\bibinfo{volume}{102}}, \bibinfo{pages}{014101}
  (\bibinfo{year}{2009}).

\bibitem{shayeganfar2009multifractal}
\bibinfo{author}{Shayeganfar, F.}, \bibinfo{author}{Jabbari-Farouji, S.},
  \bibinfo{author}{Movahed, M.~S.}, \bibinfo{author}{Jafari, G.} \&
  \bibinfo{author}{Tabar, M. R.~R.}
\newblock \bibinfo{journal}{\bibinfo{title}{Multifractal analysis of light
  scattering-intensity fluctuations}}.
\newblock {\emph{\JournalTitle{Physical Review E}}}
  \textbf{\bibinfo{volume}{80}}, \bibinfo{pages}{061126}
  (\bibinfo{year}{2009}).

\bibitem{koohi2021coupled}
\bibinfo{author}{Koohi~Lai, Z.}, \bibinfo{author}{Namaki, A.},
  \bibinfo{author}{Hosseiny, A.}, \bibinfo{author}{Jafari, G.} \&
  \bibinfo{author}{Ausloos, M.}
\newblock \bibinfo{journal}{\bibinfo{title}{Coupled criticality analysis of
  inflation and unemployment}}.
\newblock {\emph{\JournalTitle{Entropy}}} \textbf{\bibinfo{volume}{23}},
  \bibinfo{pages}{42} (\bibinfo{year}{2021}).

\bibitem{shayeganfar2010stochastic}
\bibinfo{author}{Shayeganfar, F.}, \bibinfo{author}{Jabbari-Farouji, S.},
  \bibinfo{author}{Movahed, M.~S.}, \bibinfo{author}{Jafari, G.~R.} \&
  \bibinfo{author}{Tabar, M. R.~R.}
\newblock \bibinfo{journal}{\bibinfo{title}{Stochastic qualifier of gel and
  glass transitions in laponite suspensions}}.
\newblock {\emph{\JournalTitle{Physical Review E}}}
  \textbf{\bibinfo{volume}{81}}, \bibinfo{pages}{061404}
  (\bibinfo{year}{2010}).

\bibitem{castaing1990velocity}
\bibinfo{author}{Castaing, B.}, \bibinfo{author}{Gagne, Y.} \&
  \bibinfo{author}{Hopfinger, E.}
\newblock \bibinfo{journal}{\bibinfo{title}{Velocity probability density
  functions of high reynolds number turbulence}}.
\newblock {\emph{\JournalTitle{Physica D: Nonlinear Phenomena}}}
  \textbf{\bibinfo{volume}{46}}, \bibinfo{pages}{177--200}
  (\bibinfo{year}{1990}).

\end{thebibliography}




\end{document}